\documentclass[twocolumn,twoside,showpacs,preprintnumbers,aps,prl,superscriptaddress,amsfonts,amsmath,amssymb,floatfix]{revtex4-1}

\usepackage[pdftex]{graphicx} 
\usepackage{dcolumn} 
\usepackage{bm}
\usepackage{amsmath}
\usepackage{amssymb}
\usepackage{color}
\usepackage{tikz}
\newcommand{\zeit}{\ensuremath{t}}
\newcommand{\textcelsius}{\ensuremath{^{\circ}}\textrm{C}}

\newcommand{\CapitalS}{\ensuremath{S}}

\newcommand{\Eq}[1]{Eq.~(\ref{eq:#1})}
\newcommand{\Eqs}[1]{Eqs.~(\ref{eq:#1})}
\newcommand{\eq}[1]{(\ref{eq:#1})}

\newcommand{\Fig}[1]{Fig.~\ref{fig:#1}}

\newcommand{\ie}{\emph{i.e.,}~}

\newcommand{\rmd}{\ensuremath{\mathrm{d}}}

\begin{document}

%%%%%%%%%%%%%%%%%%%%%%%%%%%%%%%%%%%%%%%%%%%%%%%%%%%%%%%%%%%%%%
\title{Breath Figures: Nucleation, Growth, Coalescence,\\ and the Size Distribution of Droplets}

\author{Johannes Blaschke}
\affiliation{Max-Planck-Institut f\"ur Dynamik und Selbstorganisation (MPIDS), 37077 G\"ottingen, Germany}
\affiliation{Fakult\"at f\"ur Physik, Universit\"at G\"ottingen, 37077 G\"ottingen, Germany}

\author{Tobias Lapp}
\affiliation{Max-Planck-Institut f\"ur Dynamik und Selbstorganisation (MPIDS), 37077 G\"ottingen, Germany}
\affiliation{Fakult\"at f\"ur Physik, Universit\"at G\"ottingen, 37077 G\"ottingen, Germany}

\author{Bj\"orn Hof}
\affiliation{Max-Planck-Institut f\"ur Dynamik und Selbstorganisation (MPIDS), 37077 G\"ottingen, Germany}

\author{J\"urgen Vollmer} 
\affiliation{Max-Planck-Institut f\"ur Dynamik und Selbstorganisation (MPIDS), 37077 G\"ottingen, Germany}
\affiliation{Fakult\"at f\"ur Physik, Universit\"at G\"ottingen, 37077 G\"ottingen, Germany}

\date{\today}

\begin{abstract}
  The analysis of the size distribution of droplets condensing on a
  substrate (breath figures) is a test ground for scaling theories.
  Here, we show that a faithful description of these distributions
  must explicitly deal with the growth mechanisms of the droplets.
  This finding establishes a gateway connecting nucleation and growth
  of the smallest droplets on surfaces to gross features of the
  evolution of the droplet size distribution.
\end{abstract}

\pacs{
05.65.+b,% 	Self-organized systems
89.75.Da,% 	Systems obeying scaling laws 
68.43.Jk% 	Diffusion of adsorbates, kinetics of coarsening and aggregation 
}

\keywords{
breath figures,
droplet deposition,
droplet merging,
self-similar droplet distribution,
fractal packings,
drop-wise condensation
}
                             
\maketitle
%%%%%%%%%%%%%%%%%%%%%%%%%%%%%%%%%%%%%%%%%%%%%%%%%%%%%%%%%%%%%%

\begin{figure*}
  \hfill
  \includegraphics[width = 0.4 \textwidth]{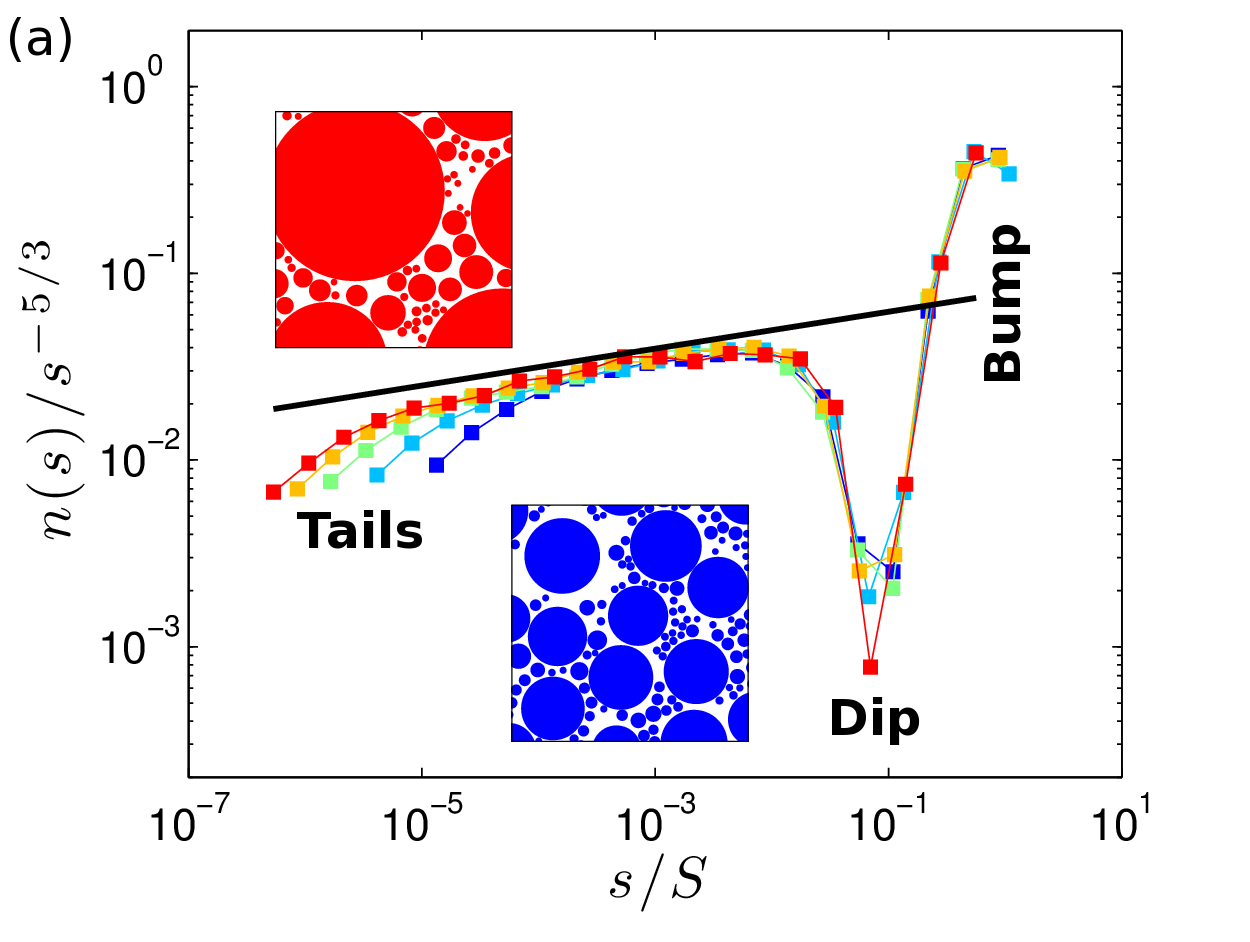} 
  \hfill
  \includegraphics[width = 0.4\textwidth]{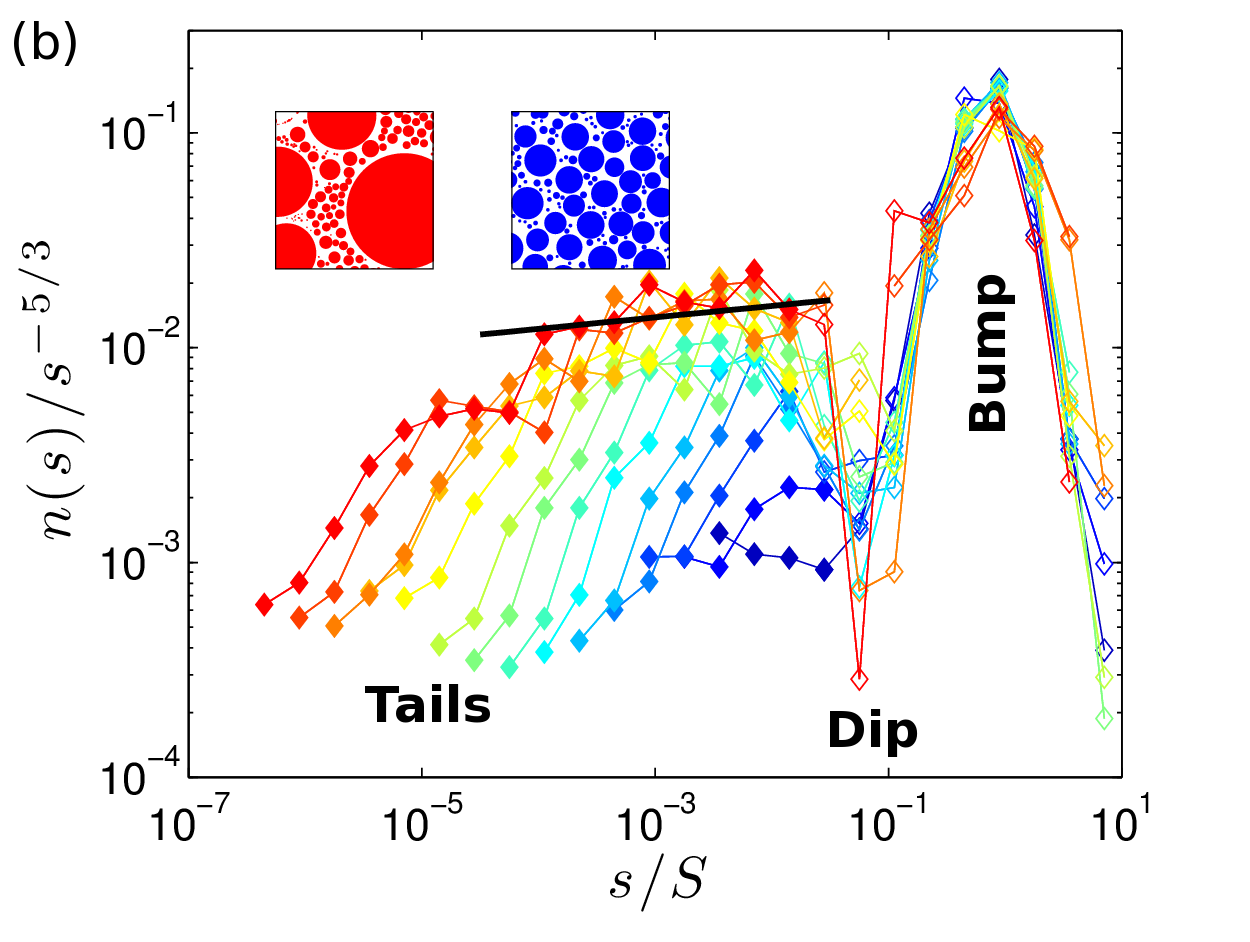}
  \hfill \ 
  \caption[]{(color online) Scaling plots of the droplet number density
    $n(s,\zeit)$ for (a) numerical and (b) experimental
    data, for (a) $8\times 10^6$ (rightmost \emph{tails}),
    $12\times 10^6$, $16\times 10^6$, $20\times 10^6$, and $24\times
    10^6$ (leftmost \emph{tails}) droplets added to a domain of size
    $1600\times 1600$ \cite{Note2}, and (b) water droplets on a
    polyethylene film \cite{Note1} where eventually the
    droplet diameters cover the range from a few microns to a few
    millimeter.
    The respective positions of the \emph{tail}, \emph{dip} and \emph{bump} of the
    distributions are indicated, and the slope, $\theta-\tau=1/12$, of the 
    scaling prediction, \Eq{tau}, is shown by solid lines. 
    The insets show snapshots of the breath figures at an early
    [blue, (lower) right] and a late  [red, left] time,
    respectively. 
    [MOVIES] of their time evolution and full details of data
    assimilation and evaluation, as well as plots of the raw data 
    are given in [SUPPLEMENTARY MATERIAL]. 
    \label{fig:nA}}
\end{figure*}

Classical questions regarding breath figures involve the influence of
material defects and impurities on the droplet patterns
\cite{Rayleigh11,Baker1922,Lopez1993,Lepopoldes2006,Sikarwar2011}.
Presently, they are used as self-assembling templates in
micro-fabrication
\cite{Boeker2004,Haupt2004,Lepopoldes2006,Wang2007,Rykaczewski2011,Samuel2011},
as highly efficient means for heat exchange in cooling systems
\cite{Mei2011,Sikarwar2011,Leach2006,Rose1973}, and they are promising
candidates for water recovery in (semi-)arid regions
\cite{Lekouch2011,Nikolayev1996}.
For these applications a detailed knowledge of the droplet size
distribution and the average droplet growth speed is vital.
Here, we demonstrate that the state of the art scaling theory
\cite{Beysens1991,Meakin1992,Viovy1988,Family1989,Blackman2000}
fails to describe data from simulations\footnote{%
  Full details on the numerical method as well as raw data
  and a [MOVIE] of the evolution are provided in the [SUPPLEMENTARY
  MATERIAL].} 
and laboratory experiments~%
\footnote{Droplet patterns are obtained from water droplets suspended
  from a polyethylene film covering a glass plate kept at a
  temperature of $20\textcelsius$ which is placed about $5\;$cm atop a
  water bath with a temperature of $60\textcelsius$. Details on the
  experimental method as well as raw data and a [MOVIE] of the evolution
  are provided in the [SUPPLEMENTARY MATERIAL].},
\Fig{nA}. A faithful description must therefore explicitly
address the microscopic growth mechanisms of droplets.

Classical scaling \cite{Beysens1991,Meakin1992} asserts that on clean
surfaces the coagulation of droplets organizes the systems into a
state where the number of droplets, $n (s, t)$, per unit droplet
volume and surface area takes a universal scaling form,
\begin{subequations}
\begin{equation} 
  n (s, \zeit) 
  =
    s^{-\theta} \; 
    f\!\left( \frac{s}{\CapitalS} \right) \, , \quad 
    \textrm{with }
    \CapitalS = \CapitalS(\zeit) \, .
  \label{eq:nA-FM} 
\end{equation}
Here $s$ denotes the droplet volume, $\theta$ is a scaling exponent,
$f(x)$ is a dimensionless function, and $\CapitalS(\zeit)$ is the
volume of the largest droplets encountered at time $\zeit$, \ie the
average volume of droplets in the \emph{bump} of the distributions
shown in \Fig{nA}.

Since $n(s,t)$ has a dimension of length to the power~$-5$ the
exponent $\theta$ must be set to a value of $\theta =5/3$
\cite{Viovy1988,Family1989,Meakin1992}.  The time evolution of $S(t)$
is found by observing that the total volume of all droplets grows
linearly in time when a constant volume flux impinges onto the
surface. In agreement with experimental and numerical observation
\cite{Viovy1988,Family1989,Meakin1992} this entails $S(t) \sim t^3$.
Moreover, a lower cutoff to the scaling at a scale $s_0/S$ has been
accounted for by a polydispersity exponent $0 < \tau < 2$
\cite{Cueille1997}. For our numerical scheme, where the mass
flux onto the surface is implemented as sustained addition of droplets
of size $s_0$ to random positions of the surface and where overlapping
droplets are subsequently merged \cite{Note2},
it was predicted~\cite{Blackman2000} to be
\begin{equation} 
  x \ll 1 \Rightarrow 
  f(x) \sim x^{\theta-\tau} \, ,  \quad  \tau =  19/12.
  \label{eq:tau}
\end{equation}%
\label{eq:scaling}%
\end{subequations}

%%%%%%%%%%%%%%%%%%%%%%%%%%%%%%%%%%%%%%%%%%%%%%%%%%%%%%%%%%%%%%
%%%%%%%%%%%%%%%%%%%%%%%%%%%%%%%%%%%%%%%%%%%%%%%%%%%%%%%%%%%%%%

The scaling, \Eq{nA-FM}, provides an excellent data collapse of the
\emph{bump} and the \emph{dip} of the numerical, \Fig{nA}(a), and the
experimental data, \Fig{nA}(b). Beyond the \emph{dip} one can discern
a self-similar scaling regime, \Eq{tau}, in the numerical data, and in
only those experimental data with the vastest range of droplet
sizes. On the other hand, in either case --- and particularly
pronounced in the experimental data --- noticeable deviations,
\emph{tails}, from the scaling prediction arise for small values of
$s/\CapitalS$.
 
In the following, we show that these deviations result from 
features of droplet growth at the small length scale, $s_0$. 
Similar to the approaches in the theory of critical phenomena
\cite{Kadanoff67,Hilfer92} or of the effect of rough boundaries in
turbulent flows \cite{Barenblatt04,Goldenfeld06}, scaling will be
recovered by asymptotic analysis \cite{BarenblattBook}, which allows us
to explicitly account for different growth mechanisms of small
droplets.
Universal and non-universal features of the asymptotic droplet density
distribution will be disentangled by discussing the consequences of the
different growth mechanisms for the small droplets in the numerical and
experimental setting, respectively.

\paragraph{Relation to fractal packings.---}%
To explore the role of the lower cutoff of scaling we consider the droplet
arrangement in breath figures as an example of a fractal packing of
disks (see \cite{Herrmann,Amirjanov:2006wn,VarratoaFoffia2012}
for recent applications in other fields), and adopt scaling arguments
developed to characterize (disordered) fractal structures to the
problem at hand: 
We assert that in the scale-separation limit, $s_0 \ll \CapitalS$, the
free surface area, \ie the area not covered by droplets, approaches a
fractal with a fractal dimension $d_f < 2$.  
Self-similarity with fractal dimension $d_f$ amounts then to the
statement that in an area of size $\CapitalS^{2/3}$ a number
\begin{equation}
  N(s_*, \CapitalS) 
  \sim \left( \frac{\CapitalS^{1/3}}{s_*^{1/3}} \right)^{d_f} 
  =  \left( \frac{s_*}{\CapitalS} \right)^{-d_f/3} 
  \label{eq:drop-number}
\end{equation}
of regions of size $s_*^{2/3}$ are required to cover the complement of
the surface area covered by all droplets larger than $s_*$. The fraction
of this area in the considered domain of size $\CapitalS^{2/3}$
amounts to
\begin{equation}
p(s_*,\CapitalS) 
= N(s_*, \CapitalS) \; \frac{s_*^{2/3}}{ \CapitalS^{2/3}}
\sim \left( \frac{s_*}{\CapitalS} \right)^{(2-d_f)/3} \, .
\label{eq:ps-star}
\end{equation}
Following \cite{Herrmann} we denote the surface area not covered by
droplets as \emph{porosity}, $p(t)$.
It is obtained by evaluating \Eq{ps-star} for the size $s_0$ 
characterizing the small scale cutoff of the fractal, 
$p(t) = p(s_0,\CapitalS(t))$.

By its definition the porosity is related the area $A_d$ covered by
droplets in a region of area $A_s$ via $p(t) = 1-A_d/A_s$: when 
the surface area in between droplets approaches a fractal of zero
measure one obtains
\[
  \int_{s_0}^{\infty} a(s) \; n(s,t) \; \rmd s 
  \equiv  \frac{ A_d }{ A_s } 
  = 1-p(t) \quad \xrightarrow{s_0/\CapitalS\to 0} \quad 1 \, ,
\]
where $a(s)$ denotes the area covered by droplets of size $s$. 
Using \Eq{scaling} with $\theta=5/3$, \Eq{tau}, $a(s)\sim s^{2/3}$,
and introducing $x=s/\CapitalS$ one obtains
\begin{eqnarray}
  p(t) 
  &=& 1 - \frac{A_d}{A_s} 
  \sim  \int_0^{s_0/\CapitalS}  x^{-1} \; f(x) \; \rmd x 
  \nonumber \\[2mm] 
  &\sim&  \int_0^{s_0/\CapitalS}  x^{\theta-\tau-1}  \; \rmd x 
  = \left( \frac{s_0}{\CapitalS(\zeit)} \right)^{\theta-\tau} \, ,
\label{eq:pt}
\end{eqnarray}
and comparing \Eqs{pt} and \eq{ps-star} yields 
\begin{equation}
  \theta - \tau = (2 - d_f)/3 \, .
  \label{eq:df}
\end{equation}
Hence, the nontrivial scaling, \Eq{tau}, of $f(x)$ for small $x$ 
reflects the fractality of the arrangements of droplets in breath
figures with a large scale separation $s_0 \ll \CapitalS$, that can
faithfully be regarded as a fractal.

This provides an independent, more accurate, means to test the
polydispersity exponent: 
For $\tau=19/12$ and $\CapitalS \sim t^3$, \Eq{pt} implies 
\( % begin{equation}
  p(t) \sim t^{-1/4} \, .
  \label{eq:pt-prediction}
\) %end{equation}
Remarkably, none of our data follow this prediction [Figs.~1(b) and
5(b) in the supplementary material]: Rather than $1/4$ we find $0.30$
for the numerical, and $0.16$ for the experimental data.

\begin{figure}
  \rule{.015\textwidth}{0mm}
  \includegraphics[width = 0.42 \textwidth]{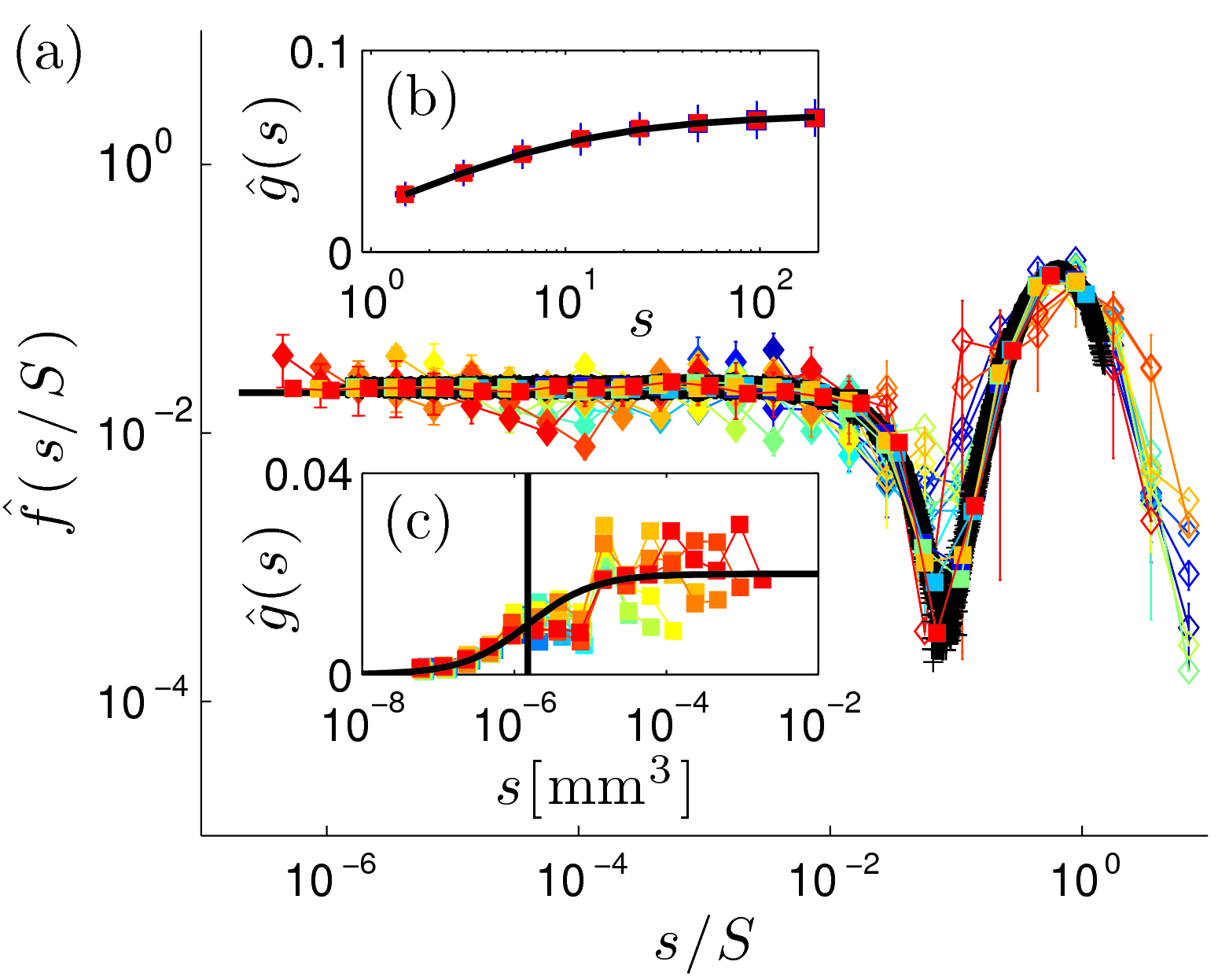}
  \rule{.015\textwidth}{0mm}
  \caption[]{(color online) 
    Master plot of the universal cutoff function, 
    $\hat f (s/\CapitalS)$, which comprises all 
    numerical and experimental data shown in \Fig{nA} (using
    the same symbols), as well as additional numerical data shown as
    black crosses (every $5 \times 10^4$ frames of $44$ runs where
    $4.23\times 10^7$ droplets are added eventually).  
    The insets show 
    $\hat g(s)$ 
    for (b) numerical and (c) experimental
    data at all times.
    Their shapes (solid lines) reflect the respective
    small-scale droplet growth mechanisms, \Eqs{dot-s-num} 
    and \eq{dot-s-exp}. 
    The vertical line in (c) marks 
    $s_0$. 
    \label{fig:multiscaling}}
\end{figure}

Hence, the different microphysics of droplet growth and merging leads
to (slightly) different fractal dimensions and a different small-scale
cutoff of scaling. To disentangle the intermediate self-similar
scaling regime from the large scale (arising from the first generation
of droplets, cf.~\cite{Fritter1991}) and the small scale physics we
introduce cutoff functions $\hat f(s/\CapitalS)$ and $\hat g(s/s_0)$
for large and small droplets, respectively:
$\hat{f}(x) \equiv f(x)/x^{\theta-\tau}$ takes a constant value 
$\hat f_0$ for
$x\ll 1$, and it accounts for the \emph{dip} and the \emph{bump} in
$n(s,t)$ for  $s \simeq \CapitalS$. 
Similarly, $\hat g(s/s_0)$ accounts for the \emph{tails} of $n(s,t)$.
As shown in the insets of \Fig{multiscaling} it approaches constant
values for $s \gg s_0$, and it takes a scaling form for all
times. 
To arrive at a complete description of the droplet size distribution
we further discuss now this lower cutoff.

%%%%%%%%%%%%%%%%%%%%%%%%%%%%%%%%%%%%%%%%%%%%%%%%%%%%%%%%%%%%%%
%%%%%%%%%%%%%%%%%%%%%%%%%%%%%%%%%%%%%%%%%%%%%%%%%%%%%%%%%%%%%%

\paragraph{Incorporating the lower cutoff.---}%
We start by writing the total volume of droplets of size $s$ per unit
volume $\mathrm{d}s$ and unit surface area 
in the form
\begin{equation}
  s \; n(s,t) 
  = 
  \CapitalS^{-2/3} \; \left( \frac{s}{\CapitalS} \right)^{-d_f/3}  \;
  \hat{f}(s/\CapitalS) \;\; \hat{g}(s/s_0) 
  \label{eq:s-nA}
\end{equation}
This expression states that in the scaling regime $s_0 \ll s \ll
\CapitalS$ the overall volume of droplets of size $s$ on an area
of size $\CapitalS^{2/3}$ is proportional to the number of droplets,
\Eq{drop-number}, of the considered size. 

In equation \eq{s-nA} the fractal dimension $d_f$ and the function 
$\hat g(s/s_0)$ are \emph{not} universal.  
We henceforth adopt the values for $d_f$ determined by fitting the
porosity, and we follow the evolution of droplets smaller
than $s_*$ over a small time interval from $t$ to $t + \rmd t$ in
order to relate the form of $\hat g(s/s_0)$ to different nucleation and
growth mechanisms of small droplets. Let the considered area on the
substrate and the time interval $\rmd t$ be chosen such that this
growth is not influenced by the merging of droplets larger than $s_*$.
When $s_*$ is so small that $\hat f(s_*/S)$ takes the constant
value $\hat f_0$, \Eq{s-nA} entails
\[ 
  s_* \; n(s_*,t) 
  = 
  \hat{f}_0 \; S^{-2/3} \; \left( \frac{s_*}{S} \right)^{-d_f/3}  \;
  \hat{g}(s_*/s_0) \, .
\]
In order to determine $\hat{g} (s_*/s_0)$ we observe that the volume density of
the droplets smaller than $s_*$ amounts to the cumulative distribution
\begin{eqnarray}
  V( s_* ) & = & \int_{s_0}^{s_*} s \; n(s,t) \; \rmd s  
  \nonumber \\[2mm]
  & = &
  \frac{\hat f_0}{S^{2/3}} \; 
  \int_{s_0}^{s_*} \left( \frac{s}{S} \right)^{-d_f/3}  \;
  \hat{g}(s/s_0)\; \rmd s \, .
  \label{eq:s-nA-star}
\end{eqnarray}
For values of $s_*$ in the scaling regime 
the increase of volume is accounted for by increasing the integral
domain.  Therefore, an infinitesimal increase of $s_*$ to 
$s_*+\dot s\;\rmd t$ in the time interval $\rmd t$ amounts to an 
increase of the volume density of droplets,
\begin{subequations}%
\begin{eqnarray}
  \frac{\rmd V}{\rmd t} 
  &=&  \frac{ V( s_* + \dot s \, \rmd t ) -  V( s_* ) }{\rmd t} \nonumber
  \\
  &=&  \frac{ \hat f_0 }{ \CapitalS^{2/3} } \; \left( \frac{s_*}{\CapitalS} \right)^{-d_f/3}  \;
  \hat{g}(s_*/s_0) \; \dot s  \, .
\end{eqnarray}
On the other hand, this change must be due to the volume flux $\Phi$
onto the fraction of area covered by droplets, 
\begin{equation}
\frac{\rmd V}{\rmd t} = p(s_*, S) \; \Phi 
\sim \left( \frac{s_*}{\CapitalS} \right)^{(2-d_f)/3} \; \Phi 
\end{equation}%
\label{eq:dV}%
\end{subequations}%
Equating the expressions for $\rmd V/\rmd t$, \Eqs{dV}, and dropping
the subscript~$*$, one hence obtains
\begin{equation}
   \hat{g}(s/s_0) \sim \frac{\Phi}{\hat f_0} \; \frac{ s^{2/3} }{ \dot s }.
   \label{eq:hat-s}
\end{equation}
This expression provides the desired connection of the average speed $\dot{s}$
of the growth of droplets of size $s$ to the form of the small-scale
cutoff $\hat{g}(s/s_0)$ of $n(s,t)$: it explains how different
microscopic droplet growth laws give rise to different
\emph{non-universal} cutoff functions $\hat{g}(s/s_0)$, and how the
universal scaling is recovered for $s \gg s_0$.
After all, the volume growth of large droplets in breath figures is
always proportional to the area exposed to the surface flux
\cite{Beysens1991,Meakin1992,Blackman2000}, $\dot s \sim \Phi \:
s^{2/3}$.
Equation \eq{hat-s} allows us to disentangle universal and
non-universal contributions to $n(s,t)$. This major finding of our
theoretical treatment is now substantiated by working out the
multiscaling predictions for the data shown in \Fig{nA}.

\paragraph{Scaling numerical data.---}%
When a small droplet, of size $s_0$, is been added to the
surface, it is merged with a droplet on the surface when the droplets
overlap. As a consequence, a droplet of radius $s^{1/3}$ will capture small
droplets of radius $s_0^{1/3}$, that are added in a distance smaller than
$s^{1/3}+s_0^{1/3}$ from its center. In the absence of other droplets, this growth
amounts to
\[
  \dot s 
  \simeq \Phi \; \left[ s^{1/3} + {s_0}^{1/3} \right]^2
  = \Phi \;  s^{2/3}  \; 
  \left[ 1 + \left(\frac{s_0}{s} \right)^{1/3} 
  \right]^2  \, .
\]
The term in square brackets accounts for an enhanced growth of small
droplets $s \gtrsim s_0$, which ceases rapidly for increasing $s$. In
practice the decay is even faster since the capture regions of
neighboring droplets overlap.
To fit the simulation data, \Fig{multiscaling}(b), one therefore
needs a non-trivial prefactor $0.76$ and an exponent close to $0.78$
rather than $1/3$,
\begin{equation}
  \hat g (s/s_0) \simeq 0.07 \; 
  \left[ 1 + 0.76 \left(\frac{s_0}{s} \right)^{0.78} 
  \right]^{-2}  \, .
  \label{eq:dot-s-num}
\end{equation}
Using \Eq{s-nA} and $\theta - \tau = 0.3$ this 
provides a perfect data collapse of all numerical data,
\Fig{multiscaling}(a).

\paragraph{Scaling experimental data.---}%
In the experimental setting, the growth rate of the droplets has two
contributions.  For small droplets, growth is limited by the
diffusion of water molecules on the substrate towards the contact line
of the droplet. As derived in \cite{Rogers1988} and observed in the
experiments of \cite{Fritter1991}, the radius of small droplets grows then
like $r \sim t^{1/4}$, such that
\(
	\dot{s} \sim \Phi \; s^{-1/3}.
\)
For larger droplets, the volume flux from the vapor phase onto the
droplets is again proportional to the exposed droplet surface, such that
\(
	\dot{s} \sim \Phi \; s^{2/3}.
\)
These growth contributions combine to
\begin{equation}
  \dot{s} 
  \sim \Phi \;  s^{2/3}
  \left( 1 + \frac{s_0}{s} \right)
  \;\;\Rightarrow\;\;
  \hat g(s/s_0) = b \; \left( 1 + \frac{s_0}{s} \right)^{-1}
  \label{eq:dot-s-exp}
\end{equation}
where $s_0 \simeq 1.5 \times 10^{-6}\; \textrm{mm}^3$ is the crossover
size scale and $b \simeq 2 \times 10^{-2}$ is a normalization
constant.  Inserting \Eq{dot-s-exp} into \Eq{hat-s} provides an
excellent prediction for $\hat{g} (s/s_0)$, \Fig{multiscaling}(c).
Also for the experimentally measured droplet
size distributions one hence obtains a perfect data collapse of the
appropriately scaled droplet number density $n(s,t)$ for all
different times, \Fig{multiscaling}(a).

\paragraph{Discussion.---}%

For the numerical data, $s_0$ amounts to the volume of the smallest
droplets in the system, \Fig{multiscaling}(b), and for the
experimental data it is about one order of magnitude larger than the
smallest observed droplets, \Fig{multiscaling}(c). In either case
$\hat g(s)$ saturates for $s \gtrsim 10^2 s_0$.
On the other hand the scaling behavior \Eq{tau} is only accessible for
values of $s$ below the \emph{dip} of the distribution, \ie for $s \lesssim
10^{-2} \CapitalS$. It can hence only be resolved in simulations where
$10^{-4} \ll s_0 / \CapitalS$, resulting in the observed scaling
regime of about $1$--$2$ decades, in the numerical data, \Fig{nA}(a),
and in the experimental data with the largest accessible scale
separation, \Fig{nA}(b).

Due to the relatively small scaling range the droplet size
distribution of breath figures can not merely be idealized as a
self-similar process with a single relevant length scale $S(t)$
\cite{Viovy1988,Family1989,Beysens1991,Meakin1992}.
Rather one explicitly has to cope with the growth law of the smallest
droplets in the system. Via its (slight) effect on the fractal
dimension characterizing the free space in between the droplets,
\Eq{df}, it sets the value of the polydispersity exponent $\tau$, and
it leads to massively different cutoff functions $\hat g(s/s_0)$,
\Fig{multiscaling}(b,c), that can completely dominate the shape of the
droplet size distribution, \Fig{nA}(b).

When both the large scale and the small scale cutoffs are properly
accounted for via \Eqs{dot-s-num} and \eq{dot-s-exp}, a remarkable
data collapse of all experimental \emph{and} numerical data into a
single plot is achieved, \Fig{multiscaling}(a).
This recovery of scaling, establishes a novel gateway connecting
features of the microscopic droplet growth on surfaces to gross
features of the evolution of the droplet size distribution.

%%%%%%%%%%%%%%%%%%%%%%%%%%%%%%%%%%%%%%%%%%%%%%%%%%%%%%%%%%%%%%
\acknowledgments

We are grateful to Bruno Eckhardt, Jens Eggers, Franziska
Gla{\ss}meier, Walter Goldburg, Siegfried Gro{\ss}mann, Andrew
Scullion, and Stephan Herminghaus for valuable discussions, and to our
referee for very valuable feedback on the manuscript.

%%%%%%%%%%%%%%%%%%%%%%%%%%%%%%%%%%%%%%%%%%%%%%%%%%%%%%%%%%%%%%

%%%%%%%%%%%%%%%%%%%%%%%%%%%%%%%%%%%%%%%%%%%%%%%%%%%%%%%%%%%%%%
%

\end{document}